 \documentclass[final,3p,times]{elsarticle}

\usepackage{graphicx}
\usepackage{amssymb}

\usepackage{amsmath,graphicx}
\usepackage{amsfonts}  
\usepackage{color}
\usepackage[keeplastbox]{flushend}
\usepackage{amssymb}
\usepackage{amsthm}
\usepackage{cite}
\usepackage{array}
\usepackage{dsfont}
\usepackage{algorithm, algorithmic}
\usepackage{bm}

\journal{Elsevier}

\begin{document}

\begin{frontmatter}


\title{\huge Second Order Statistics Analysis and Comparison between Arithmetic and Geometric Average Fusion} 


 \author[label1,label2,label3]{Tiancheng Li*}
 \author[label4]{Hongqi Fan}
 \author[label5]{Jes\'us G. Herrero}
 \author[label2,label3,label6]{Juan M Corchado}
 \address[label1]{Key Laboratory of Information Fusion Technology (Ministry of Education), School of Automation, Northwestern Polytechnical University, Xi'an 710072, China. E-mail: t.c.li@\{usal.es,mail.nwpu.edu.cn\}}
\address[label2]{BISITE research Group, University of Salamanca, Salamanca 37007, Spain}
\address[label3]{Air Institute, IoT Digital Innovation Hub, Salamanca 37188, Spain}
\address[label4]{National Key Laboratory of Science and Technology on ATR, National University of Defense Technology, Chang-Sha 410073, China. E-mail:fanhongqi@nudt.edu.cn}
\address[label5]{Department of Computer Science and Engineering, Universidad Carlos III de Madrid, Calle Madrid 126, 28903 Getafe, Spain. E-mail:jgherrer@inf.uc3m.es}
\address[label6]{Department of Electronics, Information and Communication, Osaka Institute of Technology, Osaka 535-8585, Japan}

\begin{abstract}
Two fundamental approaches to information averaging are based on linear and logarithmic combination, yielding the arithmetic average (AA) and geometric average (GA) of the fusing initials, respectively. In the context of target tracking, the two most common formats of data to be fused are random variables and probability density functions, namely 
\textit{v}-fusion and \textit{f}-fusion, respectively.
In this work, we analyze and compare the second order statistics (including variance and mean square error) of AA and GA in terms of both \textit{v}-fusion and \textit{f}-fusion. The case of weighted Gaussian mixtures representing multitarget densities in the presence of false alarms and misdetection (whose weight sums are not necessarily unit) is also considered, the result of which appears significantly different from that for a single target. In addition to exact derivation, exemplifying analysis and illustrations are provided. 
\end{abstract}


\end{frontmatter}


\section{Introduction}
\label{sec:Introduction}
The rapid development and extensive deployment of sensor/agent networks, have stemmed remarkable interest in distributed data fusion which demonstrates evident advantages in many aspects. For example, in the context of target tracking using a decentralized sensor network, the sensor cooperation can compensate for the effect of the mis-detection, false-alarms and even the failure of the local sensor and extends their fields of view, eventually resulting in improved estimation accuracy and improved robustness \citep{Chong03,hall04,Olfati-Saber04,Dimakis10,Sayed14,Campbell16}. 
Particular interest in distributed data fusion has been paid to calculating the ``average'' over the information owned by locally netted sensors/agents via peer-to-peer communication in an efficient, flexible and scalable way  \citep{Xiao04,Olfati-Saber07,Ren2013,Campbell16,Li17flooding}.  
Fundamentally, the average can be defined in two manners including, the \textit{arithmetic average} (AA) and the \textit{geometric average} (GA). Simply put, the former is a type of linear/convex fusion, akin to the linear opinion pool approach, while the latter is nonlinear/logarithmic fusion akin to the logarithmic opinion pool approach \citep{Heskes98,Hwang04}, or to say, linear versus log-linear pools \citep{Abbas09}. 
 
In the context of multi-sensor/multi-agent target tracking, the two most important types of information for fusion among local sensors/agents are
random variables (representing  parameters such as the number of targets, clutter rate, target existing probability, etc.) and probability density functions (PDFs), for which the fusion is referred to as \textit{v}-fusion and \textit{f}-fusion, respectively. While it seems that the AA fusion is more common in the former \citep{Xiao04,Olfati-Saber07,Dimakis10, Ren2013}, the GA fusion is vibrant in the latter \citep{Olfati-Saber04,Mahler00,Hlinka13}, which coincides with the Chernoff fusion \citep{Hurley02,Julier06,Nielsen11,Ahmed12} and is also known as 
covariance intersection (CI) when Gaussian functions that are uniquely characterized by the first and second order statistics are particularly considered 
\citep{Uhlmann95, Julier01, Bailey12,Chen02, Wang12,  Reinhardt15}. The CI approach was originally proposed for fusing unknown-correlated estimates produced at distinct but not necessarily independent sensors to avoid information double accounting in the fusion. Likewise, the AA fusion can also avoid information double accounting \citep{Bailey12}. 
Further approaches to combining probability  distributions of unknown cross-correlation can be found in  the literature \citep{Uhlmann03,Deng13,Tian16,Wu18,Taylor19}. In particular, a so-called generalized mean fusion approach is proposed in \citep{Taylor19} which includes both AA and GA in a unified expression. 
It is interesting to note (for discrete \textit{f}-fusion) that ``\textit{a linear opinion pool is the distribution that minimizes the sum of KL} [Kullback-Leibler] \textit{measures from the expert distributions to the aggregate distribution. A log-linear opinion pool is one that minimizes the sum of KL measures from the aggregate distribution to the expert distributions. \citep{Abbas09}}" This nice property for the GA fusion 
was earlier pointed out in \citep{Heskes98} and later extended to the PHD in \citep{Battistelli13}.

  

In addition to the posterior PDF, the fusing functions can also be the likelihood functions \citep{Hlinka12,Hlinka13,Battistelli14likelihood} or the probability hypothesis density (PHD) functions \citep{Mahler00,Clark10,Battistelli13,Uney13,Uney18,Gao19}. 
(The PHD \citep{Mahler03} differs from the PDF in that its integral over any region gives the expected number of targets in that region which can be any real number.) 
In comparison, the AA fusion has also been applied for PHD fusion 
\citep{Streit08,Streit12,Yu16,Li17gmMerging,Li17PCgm,Li18FoV,Li18CC,Li17PCsmc} %
and for raw data fusion in the means of clustering \citep{Li18fusion,Li18robust}. 
Both averaging approaches to data fusion have demonstrated, either theoretically or experimentally, gains in estimation accuracy and/or robustness, whereas deficiencies have also been identified 
from different viewpoints \citep{Heskes98,Hwang04,Clark10,Mahler09msPHD2,Mori12,Yu16,Li17PCgm,Li18FoV,Yi17, Wang17, Uney18,Taylor19}. 

Despite a few analysis about the variance alone  \citep{Julier01,Bailey12,Mahler00,Mahler09msPHD2} in \textit{f}-fusion, the mean square error (MSE) of the AA \citep{Li18CC} in \textit{v}-fusion and the divergence to the ideal fusion rule in \textit{f}-fusion \citep{Taylor19}, comprehensive analysis and comparison of their statistics  (including the variance and MSE) are still lacking. What is missing and interesting also includes the specific consideration of the multitarget density fusion in the presence of false alarms and misdetection. 

In this paper, we are not intended to investigate the motivation behind both approaches, or propose any new algorithms. Rather, we analyze and compare the statistics of the GA and of the AA in the viewpoint of point estimation, with respect to \textit{v}-fusion and \textit{f}-fusion, respectively. 
In addition to a detailed formal analysis, some approximations and exemplifying illustrations are also given
. Further, we restrict our discussion in the scalar real space $\mathbb{R}$ for simplicity, although most of the conclusions can be extended to the multidimensional case. 

The paper is organized as follows. Preliminaries are briefly introduced in Section \ref{sec:background}. Our major analysis of the \textit{v}-fusion and of the \textit{f}-fusion is given in sections \ref{sec:v-fusion} and \ref{sec:f-fusion}, respectively. A hybrid use of both approaches for averaging PHDs is discussed in Section \ref{sec:PHD-fusion}, as well as further comparison between \textit{v}-fusion and \textit{f}-fusion. Key findings are summarized by \textit{Remarks} throughout the main body of the paper and in Section \ref{sec:conclusion}.

\section{Preliminaries} \label{sec:background}
 

\subsection{Definitions}
For the unknown parameter $\theta$ of interest that takes value in a measurable region $\mathcal{\Theta} \!\subseteq\! \mathbb{R} $, the estimator $\hat{\theta}_i$ associated with PDF $f_{\hat{\mathbf{\theta}}_i}(x)$, is unbiased if it yields, on average, the true value of the parameter \citep[Section~2.3]{Kay93}, i.e., 
\begin{equation} \label{eq:unbias}
\bar{\mathbf{\theta}}_i \triangleq \mathrm{E}_{f_{\hat{\mathbf{\theta}}_i}}[\hat{\mathbf{\theta}}_i] = \int_\mathcal{\Theta} xf_{\hat{\mathbf{\theta}}_i}(x)dx = \theta \hspace{0.5mm}.
\end{equation}

The variance of the estimator $\hat{\mathbf{\theta}}_i$ is given by
\begin{equation}
\label{eq:Cov_theta}
\Sigma_{\hat{\mathbf{\theta}}_i} \triangleq  \int_\mathcal{\Theta} \big(x-\bar{\mathbf{\theta}}_i\big)^2f_{\hat{\mathbf{\theta}}_i}(x)dx = \int_\mathcal{\Theta} x^2f_{\hat{\mathbf{\theta}}_i}(x)dx - \big(\bar{\mathbf{\theta}}_i\big)^2 \hspace{0.5mm}.
\end{equation}

The MSE of an estimator $\hat{\mathbf{\theta}}_i$ is given by \citep[Section~2.4]{Kay93} \begin{equation}
\label{eq:MSE}
\mathrm{mse}(\hat{\mathbf{\theta}}_i) \triangleq  \mathrm{E}_{f_{\hat{\mathbf{\theta}}_i}(x)} [(\theta-\hat{\mathbf{\theta}}_i)^2] = \int_\mathcal{\Theta} (\theta-x)^2f_{\hat{\mathbf{\theta}}_i}(x)dx \hspace{0.5mm}. 
\end{equation}

A straightforward expansion of \eqref{eq:MSE} will lead to  
\begin{align}\label{eq:MSE_expansion}
\mathrm{mse}(\hat{\mathbf{\theta}}_i) 
= \Sigma_{\hat{\mathbf{\theta}}_i} + (\bar{\mathbf{\theta}}_i - \theta)^2 \hspace{0.5mm}.  
\end{align}
That is, the MSE of an estimator equals the sum of its variance and the square of its bias (if any). 

Furthermore, suppose that $f: X \rightarrow \mathbb{R}$ is a real-valued function whose domain is a set $X$. The set-theoretic support of $f$, denoted as $\mathrm{supp}(f)$, is the set of points in $X$ where $f$ is non-zero, i.e.,
\begin{equation} \label{eq:support}
\mathrm{supp}(f) = \{ x \in X| f(x) \neq 0 \} \hspace{0.5mm}.
\end{equation}

\subsection{Averaging in Terms of Variables: \textit{v}-fusion}
Let us consider estimators $\hat{\theta}_i, i \in \mathcal{I} \subseteq \mathbb{N}$ given in terms of random variables, such as the estimate of the number of targets \citep{Li18CC} or the target existing probability. There is no PDF or uncertainty information available and so only point estimates that are random variables are involved. 
The variable-AA is given by
\begin{equation} \label{eq:v-AA}
\hat{\theta}_v^{\mathrm{AA}} \triangleq \sum_{i\in \mathcal{I}} {\omega_i\hat{\theta}_i} \hspace{0.5mm}.
\end{equation}
Hereafter, the fusing weights are limited as $\omega_i \in (0, 1)$, $\sum_{i\in \mathcal{I}} {\omega_i}=1$. (Obviously, $\omega_i =0$ indicates that information $i$ does not really get involved in the fusion.) 

In contrast to \eqref{eq:v-AA}, the variable-GA is given by
\begin{equation} \label{eq:v-GA}
\hat{\theta}_v^{\mathrm{GA}} \triangleq \prod_{i\in \mathcal{I}} {\hat{\theta}_i}^{\omega_i} \hspace{0.5mm}.
\end{equation}
Note that, the variable-GA fusion may lead to an imaginary number when the fusing variable is negative, which is beyond the consideration of this work.  

Obviously, the GA fusion amounts to the AA fusion on the logarithms of the variables, namely (cf. \eqref{eq:v-AA}) 
\begin{equation} \label{eq:GA_log}
\log \hat{\theta}_v^{\mathrm{GA}} =\sum_{i\in \mathcal{I}} {\omega_i \log \hat{\theta}_i} \hspace{0.5mm}.
\end{equation}

\subsection{Averaging in Terms of PDFs: \textit{f}-fusion}
When the local estimator 
is given as a function such as a PDF or a PHD, the \textit{f}-fusion is involved. 
Given estimators $\hat{\theta}_i$ with PDFs  $f_{\hat{\theta}_i}(x)$, $ i \in \mathcal{I}$, the PDF-AA is given by
\begin{equation} \label{eq:AA-f}
f_{\hat{\theta}_\mathrm{AA}}(x)=\sum_{i\in \mathcal{I}} {\omega_i f_{\hat{\theta}_i}(x)} \hspace{0.5mm} ,
\end{equation}
and the PDF-GA is given by
\begin{equation} \label{eq:GA-f}
f_{\hat{\theta}_\mathrm{GA}}(x)=C^{-1} \prod_{i\in \mathcal{I}}  \big(f_{\hat{\theta}_i}(x)\big)^{\omega_i} \hspace{0.5mm},
\end{equation} 
where $C \triangleq 
\int_\mathcal{\Theta}\prod_{i\in \mathcal{I}}  \big(f_{\hat{\theta}_i}(x)\big)^{\omega_i}dx$ is a normalization term to ensure the result being a PDF (if possible). (In the generalized GA fusion applied to the PHDs \citep{Mahler00,Clark10}, such a normalization may be unnecessary. But then, the fused result of two PDFs 
may not be a PDF \citep{Clark10}. ) 

Fundamentally, the support of the AA fusion PDF is the union of those of all initial PDFs. 
In contrast, the support of $f_{\hat{\theta}_\mathrm{GA}}(x)$ is the intersection of those of all initial PDFs, 
which may be empty
. We leave here further discussion on this point and we assume all fusing estimators have the same support unless otherwise stated.

\section{Statistics Analysis for \textit{v}-Fusion} \label{sec:v-fusion}
\subsection{Bias Analysis} \label{sec:bias} 
Consider unbiased estimators $\hat{\theta}_i, i \in \mathcal{I}$ given as random variables. The unbiasedness indicates (cf. \eqref{eq:unbias})
\begin{equation} \label{eq:unbiased_fi}
\bar{\mathbf{\theta}}_i 
= \theta , \forall i \in \mathcal{I} \hspace{0.5mm}.
\end{equation}
 
Combining \eqref{eq:v-AA} and \eqref{eq:unbiased_fi} yields
\begin{align} \label{eq:unbiased_AA_v}
\bar{\theta}_v^{\mathrm{AA}} 
 = \sum_{i\in \mathcal{I}} {\omega_i}  \bar{\theta}_i  = \theta \hspace{0.5mm}.
\end{align}  
That is, the AA retains the unbiasedness (namely it remains unbiased if all fusing initials are unbiased.) 

However, as addressed the variable GA fusion may lead to an imaginary result if any $\hat{\theta}_i$ is an imaginary number and so we cannot compare it with the AA mean in general. Even in the case that $\hat{\theta}_i, \forall i \in \mathcal{I}$ are all positive, the inequality of arithmetic and geometric means \citep[Chapter 2]{steele_2004} indicates 
$\bar{\theta}_v^{\mathrm{GA}} \leq \bar{\theta}_v^{\mathrm{AA}}$ \hspace{0.5mm}. 
Here, the two means are equal if and only if ${\theta}_i = {\theta}_j, \forall i, j \in \mathcal{I}$. 
In a nutshell, \textit{the AA but not the GA retains unbiasedness in general.}


\subsection{Variance Analysis}
 \label{sec:Gau_var}
The variance of a weighted sum of multiple variables is given by the weighted sum of their covariances, \citep{Ross14book}, i.e.,
 \begin{align} \label{eq:Var_AA_sum}
     \Sigma_{\hat{\theta}_v^{\mathrm{AA}}} & = \sum_{i\in \mathcal{I}} \sum_{j\in \mathcal{I}} \mathrm{Cov}(\omega_i\hat{\theta}_i,\omega_j\hat{\theta}_j) 
     \nonumber \\ 
     & = \sum_{i\in \mathcal{I}} \omega_i^2 \Sigma_{\hat{\theta}_i} + \sum_{i<j\in \mathcal{I}}  2\omega_i\omega_j\mathrm{Cov}(\hat{\theta}_i,\hat{\theta}_j)\hspace{0.5mm}. 
 \end{align} 
Here, $\mathrm{Cov}(\hat{\theta}_1,\hat{\theta}_2)$ denotes the covariance between $\hat{\theta}_1$ and $\hat{\theta}_2$. 

Let us 
consider two variables for simplicity and define the correlation coefficient \citep[Chapt. 4]{Cohen77} as 
\begin{equation} \label{eq:rho}
    \rho \triangleq \frac{\mathrm{Cov}(\hat{\theta}_1,\hat{\theta}_2)}{ \sqrt{\Sigma_{\hat{\theta}_1}\Sigma_{\hat{\theta}_2}}} \hspace{0.5mm},
\end{equation}
taking values between $-1$ and $1$. (For two independent variables, $\rho = 0$ and for two identical variables, $\rho = 1$.) 
Then, \eqref{eq:Var_AA_sum} reduces to 
\begin{align}
\label{eq:var_theta_AA}
\Sigma_{\hat{\theta}_v^{\mathrm{AA}}} = \omega_1^2\Sigma_{\hat{\theta}_1} + \omega_2^2 \Sigma_{\hat{\theta}_2} +2\omega_1\omega_2  \rho \sqrt{\Sigma_{\hat{\theta}_1}\Sigma_{\hat{\theta}_2}} \hspace{0.5mm}.
\end{align}

We now analyze the bounds of $\Sigma_{\hat{\theta}_v^{\mathrm{AA}}}$. 
First, it is easy to verify that $ \Sigma_{\hat{\theta}_v^{\mathrm{AA}}} \leq \big(\omega_1 \sqrt{\Sigma_{1}} +\omega_2 \sqrt{\Sigma_{2}}\big)^2 \leq \mathrm{max}(\Sigma_1,\Sigma_2)$,  indicating that
\begin{equation}
\Sigma_{\hat{\theta}_v^{\mathrm{AA}}} \leq \mathrm{max}(\Sigma_1,\Sigma_2) \hspace{0.5mm},
\end{equation}
where the upper bound is approached when $w_1 \to 0$ (if $\Sigma_1\leq \Sigma_2$) or $w_1 \to 1$ (if $\Sigma_1\geq \Sigma_2$). To derive the lower bound of $\Sigma_{\hat{\theta}_v^{\mathrm{AA}}}$, we further define 
\begin{equation}
   \alpha \triangleq \frac{\Sigma_2}{ \Sigma_1} \hspace{0.5mm}. 
\end{equation}
Due to the symmetry of the expression of $\Sigma_{\hat{\theta}_v^{\mathrm{AA}}}$, we only need to consider $\alpha \geq 1$ in our analysis; the results hold by exchanging $\Sigma_2$ with $\Sigma_1$ if $\alpha < 1$. Our analysis is based on a convex function of $w \in (0,1)$ as follows 
\begin{equation} \label{eq:h(w)}
h(w;\alpha,\rho) \triangleq 1-2w+w^2+w^2\alpha+2\rho \alpha^{\frac{1}{2}}(w-w^2) \hspace{0.5mm}.
\end{equation}

Bound analysis of the function $h(w;\alpha,\rho)$ is given in Appendix A. Hereafter, we write $h(w;\alpha,\rho)$ as $h(w)$ for short. So, we have $\Sigma_{\hat{\theta}_v^{\mathrm{AA}}} = h(\omega_2)\Sigma_1$. If $\rho < \alpha^{-\frac{1}{2}}$, the fusing weights that minimize $h(w)$ are given by (cf.~\eqref{eq:aa_optimal_omega})
\begin{equation}  
\omega_1 = \frac{\alpha-\rho \alpha^{\frac{1}{2}} }{1+ \alpha-2\rho \alpha^{\frac{1}{2}}}, \hspace{2mm} \omega_2 = \frac{1 -\rho \alpha^{\frac{1}{2}} }{1+\alpha-2\rho \alpha^{\frac{1}{2}}} \hspace{0.5mm}. \label{eq:optimal-weight}
\end{equation}
This corresponds to the lower bound of $\Sigma_{\hat{\theta}_v^{\mathrm{AA}}}$ as \footnote{
In 
the case of $\rho=0$, the two fusing weights as in \eqref{eq:optimal-weight} are given by the inverse of the opposite variance divided by the sum of the two inverses. The optimal fusion reduces to the classical Millman's equation \citep{Millman40}, analogous to the weighted least squares fusion of two independent estimators.}
\begin{equation} 
\Sigma_{\hat{\theta}_v^{\mathrm{AA}}} \geq \frac{
\alpha(1-\rho^2)
}
{
1+\alpha-2\rho \alpha^{\frac{1}{2}} 
} \Sigma_1 \hspace{0.5mm}. \label{eq:vaa_lower_bounds}
\end{equation}

Otherwise (if $\rho \geq \alpha^{-\frac{1}{2}}$), the lower bound is given by 
\begin{equation} \label{eq:dual_bounds_mse_min_max}
\Sigma_{\hat{\theta}_v^{\mathrm{AA}}}  \geq \mathrm{min}(\Sigma_1,\Sigma_2) \hspace{0.5mm},
\end{equation}
where the bound is approached when $w_1 \to 1$ (if $\Sigma_1\leq \Sigma_2$) or $w_1 \to 0$ (if $\Sigma_1\geq \Sigma_2$).

\textbf{Remark 1}. \textit{For \textit{v}-fusion, the variance of the AA is upper bounded by the greatest variance of the fusing estimators. Its lower bound as given in \eqref{eq:vaa_lower_bounds} is smaller than the smallest variance of the fusing estimators if the correlation coefficient between two fusing estimator satisfies 
\begin{equation} 
    \frac{\mathrm{Cov}(\hat{\theta}_1,\hat{\theta}_2)}{ \sqrt{\Sigma_{\hat{\theta}_1}\Sigma_{\hat{\theta}_2}}}
< \bigg(\frac{\min\big(\Sigma_1,\Sigma_2\big)}{\max\big(\Sigma_1,\Sigma_2\big)} \bigg)^{\frac{1}{2}} \hspace{0.5mm}, \label{eq:min_var_AA_condition}
\end{equation}
otherwise, the lower bound is given by the smallest variance of the fusing estimators.} 

Notably, when these two variables are inversely correlated (namely $\mathrm{Cov}(\hat{\theta}_1,\hat{\theta}_2)<0$),  \eqref{eq:min_var_AA_condition} always holds.  
On the other hand, to calculate $\Sigma_{\hat{\theta}_v^{\mathrm{GA}}}$, substituting $\hat{\theta}_v^{\mathrm{AA}}, \hat{\theta}_i, i \in \mathcal{I}$ in \eqref{eq:Var_AA_sum} with $\mathrm{log}\hat{\theta}_v^{\mathrm{GA}}, \mathrm{log}\hat{\theta}_i, i \in \mathcal{I}$ (cf. \eqref{eq:GA_log}), respectively, yields 
 \begin{align} \label{eq:Var_GA_log}
     \Sigma_{\mathrm{log}\hat{\theta}_v^{\mathrm{GA}}} & = \sum_{i\in \mathcal{I}} \sum_{j\in \mathcal{I}} \mathrm{Cov}(\omega_i\mathrm{log}\hat{\theta}_i,\omega_j\mathrm{log}\hat{\theta}_j) \nonumber \\
     & = \sum_{i\in \mathcal{I}} \sum_{j\in \mathcal{I}} \omega_i\omega_j\mathrm{Cov}(\mathrm{log}\hat{\theta}_i,\mathrm{log}\hat{\theta}_j) \hspace{0.5mm}.
 \end{align} 
The above formulation involves the calculation of the covariance between (logarithmic) functions of two random variables, which can be addressed in terms of the cumulative distribution function; see, e.g., \citep{Cuadras02}. We omit further analysis on this mathematical problem, but instead, to gain insight and to illustratively compare between the AA and the GA, we study two representative examples by means of the Monte Carlo simulation. 
\subsubsection{Numerical analysis for Gaussian \textit{v}-fusion}
Note that the GA of two Gaussian variables is no longer a Gaussian variable (unless two fusing variables are identical). For numerical illustration, we here consider two approximate Gaussian distributions with $\mu_1=50, \Sigma_1=100$, and with $ \mu_2 =60, \Sigma_2=200$, respectively, in which the negative support of the Gaussian PDF (which is actually ignorable in the given examples as the negative part is far more than four-sigma to the mean of the distribution) is truncated and so all samples are guaranteed to be positively valued in order to avoid the imaginary number problem of the GA fusion.  

\begin{figure*}
\centering
\centerline{\includegraphics[width=14 cm]{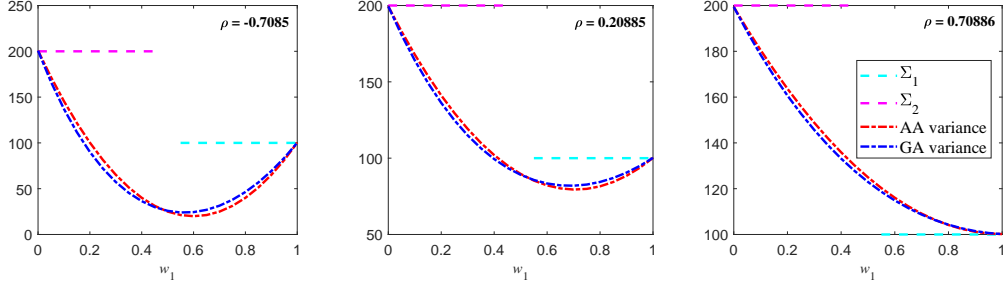}}
\caption{Variances of the AA and GA of two correlated, approximate-Gaussian-distributed variables with mean $\mu_1=50$ and variance $\Sigma_1=100$, and with mean $\mu_2 =60$ and variance $\Sigma_2=200$, respectively, under three different correlation coefficient $\rho$s. 
} \label{fig:CorGauss}
\end{figure*}

Two groups of samples are generated with different correlation coefficient $\rho$s. The corresponding values of the mean and variance of the AA and of the GA are given in Fig.~\ref{fig:CorGauss}. It is confirmed that, when $\rho =0.70846 > \alpha^{-\frac{1}{2}}$ (as $\alpha=\frac{\Sigma_2}{ \Sigma_1}=2$), we obtain dual bounds of $\Sigma_{\hat{\theta}_v^{\mathrm{AA}}}$ as in \eqref{eq:dual_bounds_mse_min_max} otherwise $\Sigma_{\hat{\theta}_v^{\mathrm{AA}}}$ can be smaller than the smallest variance of the fusing estimators. It is further seen that, 
 
\textbf{Remark 2}. \textit{The variance of the AA can be either greater or smaller than that of the GA when different fusing weights are used. There is a cross-over of their values (namely the smaller becomes the greater) as $\omega_1$ increases from 0 to 1. For certain $\rho$ and $\alpha$, the lowest AA variance that can be yielded by adjusting the fusing weights is never greater than that of the GA .}  

\subsubsection{Numerical analysis for Poisson \textit{v}-fusion}
We further consider two Poisson variables $\hat{\theta}_1\sim \mathrm{Poisson}(\lambda_1) $ and $\hat{\theta}_2 \sim \mathrm{Poisson}(\lambda_2)$, where $\lambda_1=12$ and $\lambda_2=10$ are the Poisson rates which indicate both the mean and variance of the variable. The Poisson variable is important in the tracking community, e.g., the number of targets or of false alarms that appear at a given time-interval is often modeled as a Poisson variable \citep{Mahler03,Singh09}. 
Note that both AA and GA of two Poisson variables are no longer Poisson. 

\begin{figure*}
\centering
\centerline{\includegraphics[width=14 cm]{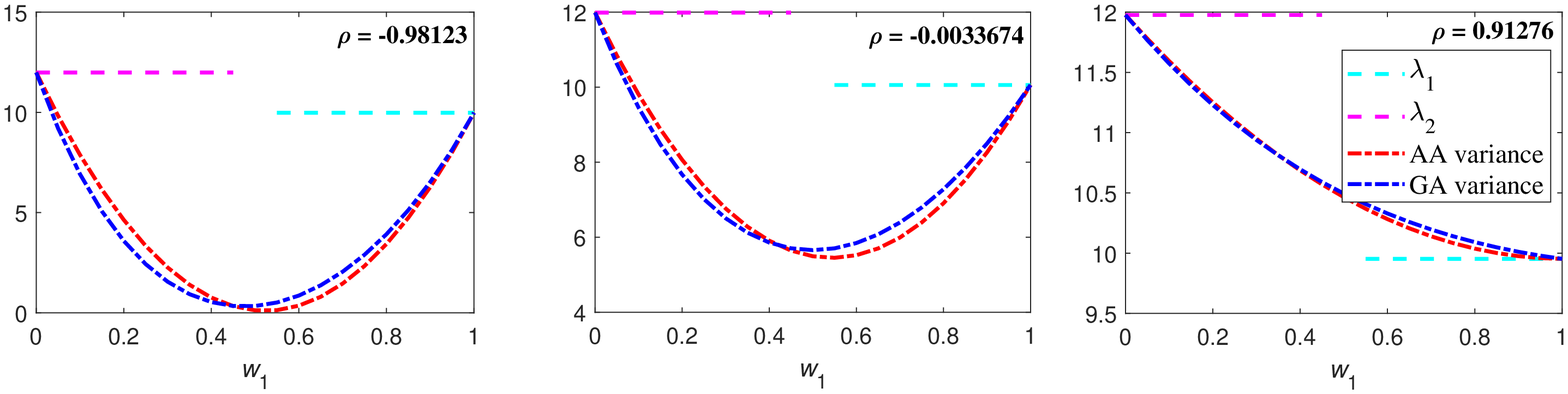}}
\caption{Variances of the AA and GA of two correlated Poisson-distributed variables with rates $\lambda_1=10,\lambda_2=12$ (and so $\alpha = \frac{\lambda_2}{\lambda_1}=1.2$), under three different correlation coefficient $\rho$s.
} \label{fig:CorPoisson}
\end{figure*}

Once more, we use the Monte Carlo method for numerical approximation. The means and variances of the AA and GA of two Poisson random variables under different correlation coefficient $\rho$s and fusing weights are given in Fig.~\ref{fig:CorPoisson}. The results are highly consistent to what shown in the Gaussian case (cf. Fig. \ref{fig:CorGauss}) and so the statement given in Remark 2 still holds.

\subsection{MSE Analysis for AA and Numerical Comparison to GA}
In this section, we study the MSE of the AA and compare it with that of the GA numerically, based on general variables that may be correlated. 

Inserting \eqref{eq:v-AA} in \eqref{eq:MSE} yields 
\begin{align}
\label{eq:mse_theta_AA_Gaussian}
\mathrm{mse}\big(\hat{\theta}_v^{\mathrm{AA}}\big) 
= & \hspace{0.5mm} \mathrm{E}_{f_{\hat{\theta}_\mathrm{AA}}(x)}
\big[\big(\omega_1(\theta-\hat{\theta}_1)+\omega_2(\theta-\hat{\theta}_2)\big)^2\big] \nonumber \\
= & \hspace{0.5mm} \omega_1^2\mathrm{mse}(\hat{\theta}_1) + \omega_2^2 \mathrm{mse}(\hat{\theta}_2) + 2\omega_1\omega_2 \beta\sqrt{\mathrm{mse}(\hat{\theta}_1)  \mathrm{mse}(\hat{\theta}_2)} \hspace{0.5mm},
\end{align}
where $\beta \triangleq \frac{\mathrm{E}_{f_{\hat{\theta}_\mathrm{AA}}(x)} \big[(\theta-\hat{\theta}_1) (\theta-\hat{\theta}_2) \big] }{\sqrt{\mathrm{mse}(\hat{\theta}_1)  \mathrm{mse}(\hat{\theta}_2)}} \in (-1,1)$.  

\begin{figure*}
\centering
\centerline{\includegraphics[width=14 cm]{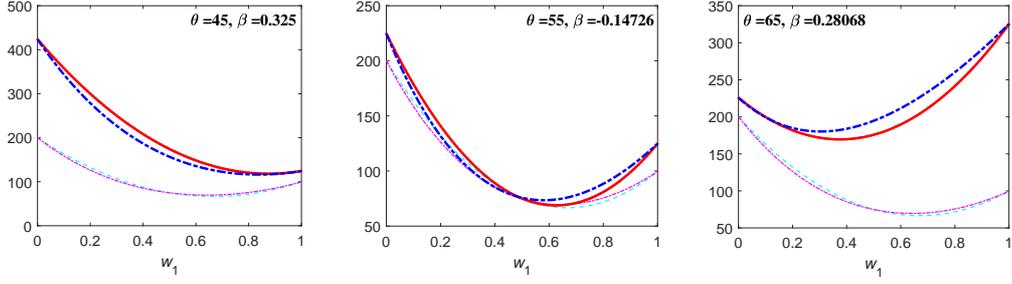}}
\caption{MSEs of the AA and GA of two independent, approximate-Gaussian-distributed variables with mean $\mu_1=50$ and variance $\Sigma_1=100$, and with mean $\mu_2 =60$ and variance $\Sigma_2=200$, respectively, in the case of three different real variables $\theta =45, 55, 65$, respectively.} \label{fig:v-mse_ind}
\end{figure*}

\begin{figure*}
\centering
\centerline{\includegraphics[width=14 cm]{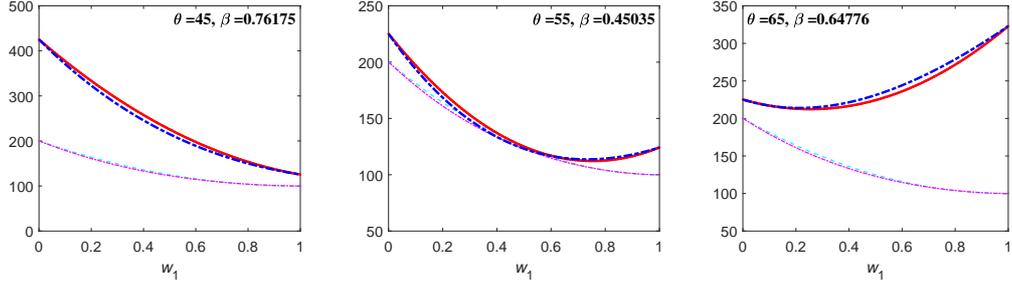}}
\caption{MSEs of the AA and GA of two correlated, approximate-Gaussian-distributed variables (with correlation coefficient $\rho=0.70736$) with mean $\mu_1=50$ and variance $\Sigma_1=100$, and with mean $\mu_2 =60$ and variance $\Sigma_2=200$, respectively, in the case of three different real variables $\theta =45, 55, 65$, respectively.} \label{fig:v-mse_cor}
\end{figure*}

As addressed, the fractional order of a Gaussian variable may involve imaginary numbers. Therefore, we cannot simply get the MSE of the GA for \textit{v}-fusion. To overcome this, once more, by means of the Monte Carlo simulation, we consider two approximate Gaussian variables $\hat{\theta}_1(x) \sim \mathcal{N}(x;50, 100)$ and $\hat{\theta}_2(x) \sim  \mathcal{N}(x;60, 200)$, for which we simulate three different real variables $\theta =45, 55$, and $65$, respectively, for different $\beta$s. The \textit{v}-fusion results are shown in Fig.~\ref{fig:v-mse_ind} for the case of two independent variables and in Fig.~\ref{fig:v-mse_cor} for the case of two correlated variable with correlation coefficient $\rho=0.70736$. It is seen that 

\textbf{Remark 3}. \textit{The MSE of the AA can be either greater or smaller than that of the GA when different fusing weights are used. The greatest discrepancy between them occurs  when the fusing weights are somewhere in the scope (0,1). The lowest bound of the MSE of either the AA or the GA is their corresponding variances, which are obtained when the AA/GA turns out to be unbiased which is only possible when the real parameter lies between two variables. Accordingly, the lower bound of the MSE of the AA is smaller than that of the GA.
} 

\subsubsection{Bounds and Comparison}
To gain analytic results on the MSE of the AA for \textit{v}-fusion, 
let us define 
\begin{equation}
    \gamma \triangleq \frac{\mathrm{mse}(\hat{\theta}_2)}{\mathrm{mse}(\hat{\theta}_1)} \hspace{0.5mm}. 
\end{equation}
Then, it can be easily varified that $\mathrm{mse}\big(\hat{\theta}_v^{\mathrm{AA}}\big) = h(\omega_2)\mathrm{mse}(\hat{\theta}_1)$, 
where $h(w)$ is defined in \eqref{eq:h(w)} (with $\rho$ and $\alpha$ replaced by $\beta$ and $\gamma$, respectively). Therefore, analogous to our analysis in Section~\ref{sec:Gau_var}, lower and upper bounds of $\mathrm{mse}\big(\hat{\theta}_v^{\mathrm{AA}}\big)$ can be obtained by using the same fusing weights $\omega_1$ and $\omega_2$ as in \eqref{eq:aa_optimal_omega}. 
Akin to Remark 1, we obtain:

\textbf{Remark 4}. \textit{For \textit{v}-fusion, the upper bound of the MSE of the AA is given by the greatest MSE of the fusing estimators. The lower bound is smaller than the smallest MSE of the fusing estimators if the correlation between two fusing estimators satisfies 
\begin{equation} \label{eq:min_mse_AA_condition}
    \beta
    < \bigg(\frac{\min\big(\mathrm{mse}(\hat{\theta}_1),\mathrm{mse}(\hat{\theta}_2)\big)}{\max\big(\mathrm{mse}(\hat{\theta}_1),\mathrm{mse}(\hat{\theta}_2)\big)} \bigg)^{\frac{1}{2}} \hspace{0.5mm}, 
\end{equation}
otherwise, the lower bound is given by the smallest MSE of the fusing estimators. } 

Notably, when the real parameter $\theta$ lies on or between $\bar{\theta}_1$ and $\bar{\theta}_2$ (namely $\mathrm{E}_\mathcal{\Theta} \big[(\theta-\hat{\theta}_1) (\theta-\hat{\theta}_2) \big] \leq 0$ and so $\beta \leq 0$), \eqref{eq:min_mse_AA_condition} always holds.

\subsubsection{Unweighted AA}
The MSE is a key metric in evaluating an  estimator/tracker \citep{Kay93}. However, in practice, $\gamma$ is often unknown since the MSE of each fusing estimator that is calculated based on the real parameter is practically unknown. \footnote{
In the literature, e.g., \citep{Julier01,Chen02,Uney13,Wang12,Reinhardt15,Li17PCgm}, the most common approach to designing the fusing weights is based on minimizing the (trace or determinant of) variance of the fused estimator, which only equals the MSE when the estimator is unbiased. However, the GA does not guarantee unbiasedness as addressed. Such a minimum variance criterion for selecting the weights can be dated at latest back to \citep{Bates69}.
} 
One may simply choose to use uniform fusing weights $\omega_1=\omega_2=\frac{1}{2}$ (namely unweighted averaging). Then, we obtain (cf.~\eqref{eq:h(w)})
\begin{equation}
h\Big(\frac{1}{2}\Big)  = \frac{1+\gamma +2\beta \gamma^{\frac{1}{2}}}{4} \hspace{0.5mm}.
\end{equation}

In this case, a sufficient and necessary condition for the un-weighted AA fusion to be ``better'' than the best fusing estimator in the sense of obtaining smaller MSE (i.e., $h\big(\frac{1}{2} \big)  < 1 $) is given by (if possible)
\begin{equation}
\beta < \frac{3-\gamma}{2 \gamma^{\frac{1}{2}}} \triangleq g(\gamma). 
\end{equation}

Calculating the derivative of $g(\gamma)$ with respect to $\gamma$ yields
\begin{align} 
\frac{\mathrm{d} \hspace{0.5mm} g(\gamma)}{\mathrm{d}\gamma} 
 = \frac{- 1 -3 \gamma^{-1}}{4\gamma^{\frac{1}{2}}} < 0 \hspace{0.5mm},   
\end{align}
which indicates that $g(\gamma)$ decreases with the increase of $\gamma$, and therefore, $g(\gamma) < -1, \forall \gamma> 9$. Since $-1 <\beta$, we therefore assert that $h\big(\frac{1}{2} \big)  < 1 $ is impossible if $\gamma> 9$. In this case, the MSE of the unweighted AA must lie between the best and the worst of the MSEs of the fusing estimators.

\section{Statistics Analysis for \textit{f}-Fusion} \label{sec:f-fusion}

In the context of Bayesian estimation, the local estimate is given in the form of a posterior PDF or PHD which contains the complete information about the distribution of the state of the interest. Then, the \textit{f}-fusion as given in \eqref{eq:AA-f} and \eqref{eq:GA-f} is needed, which as to be shown in this section is very different from the \textit{v}-fusion. 

\subsection{Bias Analysis}
\label{sec:f-fusion-bias}
Let us consider again a set of unbiased fusing estimators with PDFs as given in \eqref{eq:unbiased_fi}, we obtain from  \eqref{eq:AA-f} 
\begin{align} \label{eq:unbiased_AA_f}
\bar{\theta}_f^{\mathrm{AA}} \triangleq \mathrm{E}_{f_{\hat{\theta}_\mathrm{AA}}(x)}[\hat{\theta}_f^{\mathrm{AA}}] 
& = \int_\mathcal{\Theta} x\sum_{i\in \mathcal{I}} {\omega_i f_{\hat{\theta}_i}(x) dx} 
= \theta \hspace{0.5mm}. 
\end{align}  
That is, the AA retains the unbiasedness in the \textit{f}-fusion. In contrast, 
\begin{align} \label{eq:biased_GA_f}
\bar{\theta}_{{f\mathrm{GA}}} & \triangleq \mathrm{E}_{f_{\hat{\theta}_\mathrm{GA}}(x)}[\hat{\theta}_f^{\mathrm{GA}}] 
= \int_\mathcal{\Theta} xC^{-1}\prod_{i\in \mathcal{I}}  \big(f_{\hat{\theta}_i}(x)\big)^{\omega_i}dx \hspace{0.5mm}, 
\end{align}
which does not equal $\theta$ in general. 

In particular, \eqref{eq:biased_GA_f} does not equal $\theta$ if any $f_{\hat{\theta}_i}(x),{i\in \mathcal{I}}$ is asymmetric (which is common when the function is represented by a mixture/sum of weighted sub-functions, such as the widely used Gaussian mixture (GM)). We demonstrate this by a simple example in Appendix B. (We must note that, the above bias analysis rooted in the classic point estimation is different from the  
interpretation of the unbiasedness for an estimator in the Bayesian view \footnote{In the Bayesian formulation, the real parameter $\theta$ is considered random and the Bayesian posterior is given in the manner of  an estimate to the true distribution of $\theta$; so the unbiasedness definition as in \eqref{eq:unbias} does not directly apply to the Bayesian estimator in general}; see, e.g., \citep{Noorbaloochi18}
. In the Bayesian view, one care about the quality of the \textit{distribution} rather than a single point.
)

We add that the estimate bias can greatly reduce the probability for the fusion/filter to benefit \citep{Li16Effectiveness} in time series estimation since the bias is supposed to progagrate over time. 

\subsection{Variance Analysis (for Two Gaussian PDFs)}
\label{sec:ErrorCov}
In this section, we analyze the variances of the PDF-AA $f_{\hat{\theta}_\mathrm{AA}}(x)$ and PDF-GA $f_{\hat{\theta}_\mathrm{GA}}(x)$ for fusing two Gaussian PDFs $f_{\hat{\theta}_1}(x) = \mathcal{N}(x;\mu_{1},\Sigma_{1})$ and $f_{\hat{\theta}_2}(x) = \mathcal{N}(x;\mu_{2},\Sigma_{2})$. 
 
\subsubsection{General Result}
In the addressed case, \eqref{eq:AA-f} reduces to a GM-PDF $f_{\hat{\theta}_f^{\mathrm{AA}}}(x) = \omega_1\mathcal{N}(x;\mu_{1},\Sigma_{1}) + \omega_2\mathcal{N}(x;\mu_{2},\Sigma_{2})$ whose mean $\bar{\theta}_f^{\mathrm{AA}}$ and variance $\Sigma_{\hat{\theta}_f^{\mathrm{AA}}}$ are 
\begin{align}
 \bar{\theta}_f^{\mathrm{AA}} & = \omega_1\mu_{1}+\omega_2\mu_{2} \hspace{0.5mm}, \label{eq:Gau_AA_mean_PDF} \\ 
\Sigma_{\hat{\theta}_f^{\mathrm{AA}}} & = \omega_1\Sigma_{1} +\omega_2\Sigma_{2} + \Delta(\omega_1, \omega_2) \hspace{0.5mm},\label{eq:Gau_AA_Cov_PDF}
\end{align}
respectively, where $\Delta(\omega_1,\omega_2) \triangleq \omega_1\omega_2(\mu_{1} - \mu_{2})^2 \geq 0$. 

In contrast, the GA of two Gaussian PDFs remains a Gaussian PDF. That is, \eqref{eq:GA-f} reduces to a single Gaussian PDF $f_{\bar{\theta}_f^{\mathrm{GA}}}(x) = \mathcal{N}(x;\bar{\theta}_f^{\mathrm{GA}},\Sigma_{\hat{\theta}_f^{\mathrm{GA}}})$ with \citep{Mahler00}
\begin{align}
\Sigma_{\hat{\theta}_f^{\mathrm{GA}}} 
& =\frac{\Sigma_{1}\Sigma_{2}}{\omega_1\Sigma_{2}+\omega_2\Sigma_{1}},\label{eq:Gau_GA_Cov_PDF} \\
 \bar{\theta}_f^{\mathrm{GA}} 
 & = \frac{\omega_1\Sigma_{1}^{-1}\mu_{1}+\omega_2\Sigma_{2}^{-1}\mu_{2}}{\omega_1\Sigma_{1}^{-1}+\omega_2\Sigma_{2}^{-1}} \hspace{0.5mm}. \label{eq:Gau_GA_mean_PDF} 
\end{align}

As shown, both the mean of the AA as in \eqref{eq:Gau_AA_mean_PDF} and the mean of the GA as in \eqref{eq:Gau_GA_mean_PDF} show a linear combination of the means of the fusing estimators
. (In this sense, the CI approach is also considered as a linear fusion of  \textit{estimators} \citep{Chen02,Wang12,Reinhardt15,Wu18}.) Differently, the variances of the fusing estimators are also involved in the latter but not in the former.
In what follows, we analyze and compare their variances as in \eqref{eq:Gau_AA_Cov_PDF} and \eqref{eq:Gau_GA_Cov_PDF}.

\subsubsection{Bounds and Comparison} \label{sec:bound_var}
Given $0< \omega_1,\omega_2 < 1$, we obtain the dual, tight bounds on $\Sigma_{\hat{\theta}_f^{\mathrm{GA}}} $ from \eqref{eq:Gau_GA_Cov_PDF} 
\begin{equation} \label{eq:GA_Cov_Bound}
\min\big(\Sigma_{1}, \Sigma_{2}\big) \leq \Sigma_{\hat{\theta}_f^{\mathrm{GA}}} \leq \max\big(\Sigma_{1}, \Sigma_{2}\big) \hspace{0.5mm}, 
\end{equation}
where the equations hold when and only when $\Sigma_{1}= \Sigma_{2}$ for which $\Sigma_{\hat{\theta}_f^{\mathrm{GA}}} = \Sigma_{1}= \Sigma_{2}$, regardless of the fusing weights. Otherwise, if $\Sigma_{1} \neq \Sigma_{2}$, the bounds are approached when $\omega_1 \to 0, \omega_2 \to 1$ (for one of the dual bounds
) or when $\omega_1 \to 1,\omega_2 \to 0$ (for the other bound).  

Since $\Delta(\omega_1,\omega_2) \geq 0 $, we obtain
\begin{equation} \label{eq:AA_Cov_Bound}
\Sigma_{\hat{\theta}_\mathrm{AA}} > \omega_1\Sigma_{1} +\omega_2\Sigma_{2} \triangleq \mathrm{LB} (\Sigma_{\hat{\theta}_f^{\mathrm{AA}}}) \hspace{0.5mm}, 
\end{equation}
where, 
$\mathrm{LB} (\Sigma_{\hat{\theta}_f^{\mathrm{AA}}})$ 
is further dually, tightly bounded as (cf. \eqref{eq:GA_Cov_Bound}): 
\begin{equation}
    \min\big(\Sigma_{1}, \Sigma_{2}\big) \leq \mathrm{LB}(\Sigma_{\hat{\theta}_f^{\mathrm{AA}}}) \leq \max\big(\Sigma_{1}, \Sigma_{2}\big) \hspace{0.5mm}.
\end{equation}
 
As shown, $\Sigma_{\hat{\theta}_f^{\mathrm{AA}}}$ can not be upper bounded by the variances alone of the fusion estimators, as it also depends on the discrepancy between the means of two fusing estimators. It is hard to say whether this is an advantage or disadvantage just as whether two distant estimators should be fused or not. 
Finally, we have the following derivation 
\begin{align} \label{eq:Cov_AA_GA_Comp}
\mathrm{LB}(\Sigma_{\hat{\theta}_f^{\mathrm{AA}}})
& = \frac{\big(\omega_1\Sigma_{1} +\omega_2\Sigma_{2}\big)\big(\omega_1\Sigma_{2}+\omega_2\Sigma_{1}\big) }{\omega_1\Sigma_{2}+\omega_2\Sigma_{1}} \nonumber \\
& = \frac{(\omega_1^2+\omega_2^2)\Sigma_{1}\Sigma_{2}+  \omega_1\omega_2(\Sigma_{1}^2+\Sigma_{2}^2)}{\omega_1\Sigma_{2}+\omega_2\Sigma_{1}} \nonumber \\
& \geq \frac{(\omega_1^2+\omega_2^2)\Sigma_{1}\Sigma_{2}+ 2 \omega_1\omega_2(\Sigma_{1} \Sigma_{2} )}{\omega_1\Sigma_{2}+\omega_2\Sigma_{1}} \nonumber \\
&  = \frac{\Sigma_{1}\Sigma_{2}}{\omega_1\Sigma_{2}+\omega_2\Sigma_{1}}= \Sigma_{\hat{\theta}_f^{\mathrm{GA}}}\hspace{0.5mm}. 
\end{align}

\begin{figure*}
\centering
\centerline{\includegraphics[width=15 cm]{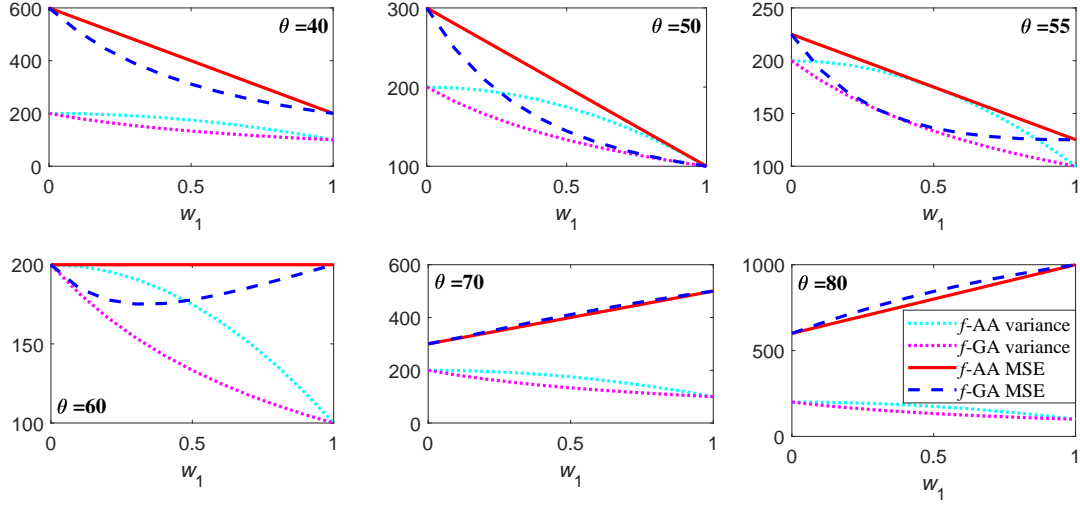}}
\caption{Variances and MSEs of the AA and of the GA, of two Gaussian PDFs $f_{\hat{\theta}_1}(x) = \mathcal{N}(x;50, 100)$ and $f_{\hat{\theta}_2}(x) = \mathcal{N}(x; 60, 200)$ regarding different real variables $\theta \in [40, 80]$ and different fusing weights $w_1 \in [0, 1]$.} \label{fig:f-MSE_100_200}
\end{figure*}

In summary, we have the following remark (cf. Remark 2 for \textit{v}-fusion). Numerical demonstration will be given in Fig.~\ref{fig:f-MSE_100_200} later on. Inflated variance due to AA or GA has also been pointed out in part by \citep{Mahler09msPHD2,Bailey12}, etc. 

\textbf{Remark 5}. \textit{For Gaussian \textit{f}-fusion, the AA fusion always leads to a greater variance than the GA fusion does when they use the same fusing weights while the variance of the GA, but not that of the AA, is bounded by the smallest and greatest variances of the fusing estimators. }

\begin{figure*}
\centering
\centerline{\includegraphics[width=15 cm]{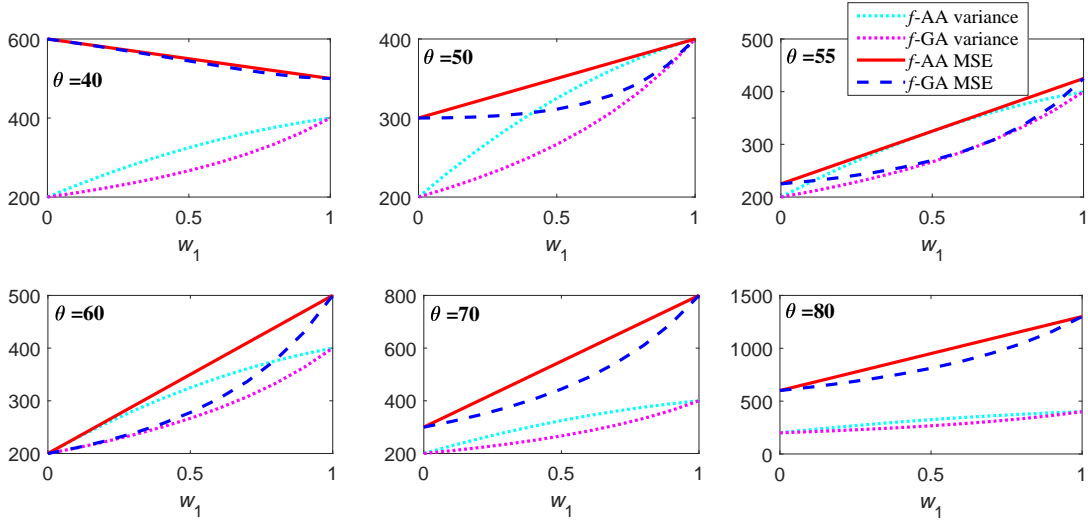}}
\caption{Variances and MSEs of the AA and of the GA, of two Gaussian PDFs $f_{\hat{\theta}_1}(x) = \mathcal{N}(x;50, 400)$ and $f_{\hat{\theta}_2}(x) = \mathcal{N}(x;60, 200)$ regarding different real variables $\theta \in [40, 80]$ and different fusing weights $w_1 \in [0, 1]$.} \label{fig:f-MSE_400_200}
\end{figure*}

\subsection{MSE Analysis} \label{sec:mse}
The MSE of $\hat{\theta}_f^{\mathrm{AA}}$ 
is calculated by (cf.~\eqref{eq:MSE})
\begin{align}
\label{eq:mse_theta_AA_f}
\mathrm{mse}\big(\hat{\theta}_f^{\mathrm{AA}}\big) 
& = \int_\mathcal{\Theta} (\theta-x)^2 \sum_{i\in \mathcal{I}} {\omega_i f_{\hat{\theta}_i}(x)} dx \nonumber \\
& = \sum_{i\in \mathcal{I}} \omega_i \int_\mathcal{\Theta} (\theta-x)^2 {f_{\hat{\theta}_i}(x)} dx \nonumber \\
& = \sum_{i\in \mathcal{I}} \omega_i \mathrm{mse}\big(f_{\hat{\theta}_i}(x)\big) \hspace{0.5mm} 
\end{align}
which simply indicates that (cf. Remark 4 for \textit{v}-fusion):

\textbf{Remark 6}. \textit{The AA has an MSE that is the linearly weighted average of the MSEs of the fusing estimators and the MSE of the AA is bounded by the smallest and greatest MSEs of the fusing estimators.}

Expression \eqref{eq:MSE_expansion} provides an easy way to calculate the MSE of $\hat{\theta}_f^{\mathrm{GA}}$ based on \eqref{eq:Gau_GA_Cov_PDF} and \eqref{eq:Gau_GA_mean_PDF}, i.e., 
\begin{align}
\label{eq:mse_theta_GA_f}
\mathrm{mse}\big(\hat{\theta}_f^{\mathrm{GA}}\big) 
 &= \Sigma_{\hat{\theta}_f^{\mathrm{GA}}} + \big( \bar{\theta}_f^{\mathrm{GA}} - \theta \big)^2 \nonumber \\
& = \frac{\Sigma_{1}\Sigma_{2}}{\omega_1\Sigma_{2}+\omega_2\Sigma_{1}} + \bigg( \frac{\omega_1\Sigma_{1}^{-1}\mu_{1}+\omega_2\Sigma_{2}^{-1}\mu_{2}}{\omega_1\Sigma_{1}^{-1}+\omega_2\Sigma_{2}^{-1}} - \theta \bigg)^2
\nonumber \\
& = \underbrace{\frac{\Sigma_{1}\Sigma_{2}}{\omega_1\Sigma_{2}+\omega_2\Sigma_{1}}}_{\triangleq \mathrm{mse}_1\big(\hat{\theta}_f^{\mathrm{GA}}\big)} + \underbrace{\big( a \xi_1 +b \xi_2 \big)^2}_{\triangleq \mathrm{mse}_2\big(\hat{\theta}_f^{\mathrm{GA}}\big)}\hspace{0.5mm}. 
\end{align}
where $a \triangleq \frac{\omega_1\Sigma_{1}^{-1}}{\omega_1\Sigma_{1}^{-1}+\omega_2\Sigma_{2}^{-1}}\in (0,1)$, $b \triangleq \frac{\omega_2\Sigma_{2}^{-1}}{\omega_1\Sigma_{1}^{-1}+\omega_2\Sigma_{2}^{-1}} = 1-a \in (0,1)$, $\xi_1 \triangleq \mu_{1} - \theta$ and $\xi_2\triangleq \mu_{2} - \theta$. 

It is easy to be verified that $\mathrm{mse}_1\big(\hat{\theta}_f^{\mathrm{GA}}\big) \geq \mathrm{min}(\Sigma_{1}, \Sigma_{2})$, $\mathrm{mse}_2\big(\hat{\theta}_f^{\mathrm{GA}}\big) \geq 0$, and so 
\begin{equation}
    \mathrm{mse}\big(\hat{\theta}_f^{\mathrm{GA}}\big) \geq  \mathrm{min}(\Sigma_{1}, \Sigma_{2})\hspace{0.5mm},
\end{equation}
where the equation holds when and only when both fusing Gaussian PDFs are unbiased and identical. 

We now compare between $\mathrm{mse}\big(\hat{\theta}_f^{\mathrm{GA}}\big)$ and $\mathrm{mse}\big(\hat{\theta}_f^{\mathrm{AA}}\big)$. 
First, 
combining \eqref{eq:mse_theta_AA_f} and \eqref{eq:MSE_expansion} yields 
\begin{align} \label{eq:mse_theta_AA_f_2} 
\mathrm{mse}\big(\hat{\theta}_f^{\mathrm{AA}}\big) =  \underbrace{\omega_1 \Sigma_{1} + \omega_2\Sigma_{2} }_{\triangleq \mathrm{mse}_1\big(\hat{\theta}_f^{\mathrm{AA}}\big)} + \underbrace{\omega_1\xi_1^2 + \omega_2\xi_2^2}_{\triangleq \mathrm{mse}_2\big(\hat{\theta}_f^{\mathrm{AA}}\big)}\hspace{0.5mm}.
\end{align}

We have the following simple derivation
\begin{align}
\big( \omega_1 \Sigma_{1} + \omega_2\Sigma_{2}\big) \big( \omega_2 \Sigma_{1} + \omega_1\Sigma_{2}\big)  
& = 
\big(\omega_1^2+ \omega_2^2\big) \Sigma_{1} \Sigma_{2} + \omega_1\omega_2\big( \Sigma_{1}^2+ \Sigma_{2}^2\big)
\nonumber \\
& \geq \big(\omega_1^2+ \omega_2^2\big) \Sigma_{1} \Sigma_{2} + 2\omega_1\omega_2\big( \Sigma_{1} \Sigma_{2} \big)
= \Sigma_{1} \Sigma_{2} \hspace{0.5mm},
\end{align}
which indicates that (as long as both averaging approaches use the same fusing weights)
\begin{equation} \label{eq:mse_1}
\mathrm{mse}_1\big(\hat{\theta}_f^{\mathrm{GA}}\big) \leq \mathrm{mse}_1\big(\hat{\theta}_f^{\mathrm{AA}}\big) \hspace{0.5mm}.
\end{equation}

To compare between $\mathrm{mse}_2\big(\hat{\theta}_f^{\mathrm{GA}}\big)$ and $\mathrm{mse}_2\big(\hat{\theta}_f^{\mathrm{AA}}\big)$, we consider two specific cases: First, if both fusing Gaussian PDFs are unbiased, i.e., $\mu_{1} = \mu_{2} =\theta$, we have $\mathrm{mse}_2\big(\hat{\theta}_f^{\mathrm{GA}}\big) =\mathrm{mse}_2\big(\hat{\theta}_f^{\mathrm{AA}}\big)$ and further by using \eqref{eq:mse_1}, 
\begin{equation}
    \mathrm{min}\big(\Sigma_{1},\Sigma_{2} \big) \leq \mathrm{mse}\big(\hat{\theta}_f^{\mathrm{GA}}\big)\leq \mathrm{mse}\big(\hat{\theta}_f^{\mathrm{AA}}\big)\leq \mathrm{max}\big(\Sigma_{1},\Sigma_{2} \big) \hspace{0.5mm},  
\end{equation}
where the bounds are approached when the two fusing weights approach $0$ and $1$, respectively.
 
Secondly, if $\Sigma_{1} = \Sigma_{2}$, we have $\omega_1=a,\omega_2=b$. Subsequently, the following straightforward derivation is obtained
\begin{equation}
\mathrm{mse}_2\big(\hat{\theta}_f^{\mathrm{GA}}\big)- \mathrm{mse}_2\big(\hat{\theta}_f^{\mathrm{AA}}\big)  = - \omega_1\omega_2\big( \xi_1 - \xi_2  \big)^2 \leq 0 \hspace{0.5mm}, 
\end{equation}
namely $\mathrm{mse}_2\big(\hat{\theta}_f^{\mathrm{GA}}\big) \leq \mathrm{mse}_2\big(\hat{\theta}_f^{\mathrm{AA}}\big)$, as long as they use the same fusing weights. Combining this with \eqref{eq:mse_1} yields
\begin{equation}
    \mathrm{mse}\big(\hat{\theta}_f^{\mathrm{GA}}\big) \leq \mathrm{mse}\big(\hat{\theta}_f^{\mathrm{AA}}\big) \hspace{0.5mm}.
\end{equation}
 
\textbf{Remark 7}. \textit{If both fusing Gaussian PDFs are unbiased or if they have the same variance, the MSE of the GA is smaller than or equals that of the AA and is always greater than the smallest variance of the fusing estimators. }

To gain further insight into their difference in the general case, by means of the Monte Carlo simulation, we consider two Gaussian PDFs $f_{\hat{\theta}_1}(x) = \mathcal{N}(x; \mu_1 = 50, 100)$ and $f_{\hat{\theta}_2}(x) = \mathcal{N}(x;\mu_2 =60, 200)$ and two Gaussian PDFs $f_{\hat{\theta}_1}(x) = \mathcal{N}(x;\mu_1 = 50, 400)$ and $f_{\hat{\theta}_2}(x) = \mathcal{N}(x;\mu_2 = 60, 200)$, respectively. The results are shown in Fig.~\ref{fig:f-MSE_100_200} and Fig.~\ref{fig:f-MSE_400_200}, respectively, for the real parameter $\theta \in [40, 80]$ 
and fusing weight $\omega_1 \in (0, 1)$. It is seen that (cf. Remark 3 for \textit{v}-fusion; relevant results can be found in \citep{Taylor19})

\textbf{Remark 8}. \textit{For Gaussian \textit{f}-fusion, the MSE of the AA 
is in most cases greater than that of the GA, unless $\theta$ is considerably greater than $\mathrm{max}(\mu_1,\mu_2)$ and the fusing estimator that has a greater mean has a greater variance. Different from the case of \textit{v}-fusion, there is no cross-over of their MSEs when the fusing weights change. That is, for certain PDFs and real parameter, one is always better than the other. } 


\section{PHD Averaging} \label{sec:PHD-fusion}
When multiple objects are involved, the fusing distributions are ``multimodal''such as typically a mixture of sub-functions (each of which can be referred to as a ``component'') like a GM whose integral is no more (but usually greater than) unit. More precisely, finite set distributions \citep[Ch. 5]{Daley02}  such as the PHD \citep{Mahler03} 
factories into a cardinality distribution on the number of objects and a localization density conditioned on the cardinality. 
In this case, while the AA of a sum can be straightforwardly expressed as a cascaded sum of the fusing sums (after re-weighting them) that remains in the same form \citep{Li17gmMerging,Li17PCgm}te, the fractional order exponential power of a sum 
does not remain as a sum of the same form, and typically approximation must be resorted to; see, e.g., \citep{Mariam07,Ahmed12,Battistelli13,Gunay16,Li18Nehorai}. 

\subsection{Approximate GM-GA Fusion}

By omitting the cross-products of different Gaussian functions/components (GCs -- we note here that, this simplification only make sense in the case where the GCs in the mixture are well distant), 
 the fractional order exponential power of a mixture of $n$ GCs 
 can be approximated by 
\begin{equation}\label{eq:GM_fractional_power}
\bigg(\sum_{i=1}^{n} w_i \mathcal{N}({x};{m}_i,{P}_i)\bigg)^\omega \approx \sum_{i=1}^{n} \big(w_i \mathcal{N}({x};{m}_i,{P}_i)\big)^\omega \hspace{0.5mm},
\end{equation}
where the covariance inflation of CI, for a weighted Gaussian PDF is equivalent to raising the Gaussian function to a power, which remains Gaussian, namely
\begin{equation}\label{eq:GC_power}
\big(w \mathcal{N}({x;m,P})\big)^\omega = w^\omega \epsilon(\omega, {P}) \mathcal{N}({x};{m}, {\omega}^{-1}{P} ) \hspace{0.5mm},
\end{equation}
where $\epsilon(\omega, {P}) =\sqrt[]{\frac{\det (2 \pi {P}\omega^{-1})}{[\det (2 \pi {P})]^\omega}} = \sqrt[]{(2 \pi {P})^{(1-\omega)}\omega^{-1}} $ \citep{Mahler00}.

In addition, the product of two GCs remains a GC, i.e.,
\begin{equation}\label{eq:GC_product}
w_1 \mathcal{N}({x};{m}_1,{P}_1)w_2 \mathcal{N}({x};{m}_2,{P}_2) = w_{12} \mathcal{N}({x};{m}_{12},{P}_{12}) \hspace{0.5mm},
\end{equation}
where ${P}_{12} = ({P}_1^{-1}+{P}_2^{-1})^{-1},{m}_{12} = {P}_{12} ({m}_1{P}_1^{-1}+{m}_2{P}_2^{-1}),w_{12}=w_1w_2\mathcal{N}({m}_1-{m}_2;0,{P}_1 + {P}_2)$ in which the coefficient $\mathcal{N}({m}_1-{m}_2;0,{P}_1 + {P}_2)$ measures the separation of the two GCs.

By using \eqref{eq:GM_fractional_power}, \eqref{eq:GC_power} and \eqref{eq:GC_product}, the approximate GA-fusion of two GMs is ready to be obtained; the interested readers are kindly referred to \citep{Battistelli13} for the detail. 
Two points are worth noting. First, the GM-GA fusion requires fusing all pairs of GCs between neighboring sensors, which will result in a multiplied number of GCs. That is, the GA of $J_1$ GCs and $J_2$ GCs is a mixture of $(J_1\cdot J_2)$ GCs while it is a mixture of $(J_1+ J_2)$ GCs in the case of AA fusion. We note here that, a potential means to ameliorate both averaging approaches is to apply mixture merging and pruning to reduce the number of GCs/peaks. Second, to perform the GA fusion as addressed above, the local GM-PHD needs to be normalized to a PDF (cf. \eqref{eq:GM_fractional_power}). At the end, the resultant GCs need to be properly weighted, such that their sum equals the average of the original weight sums of the fusing GMs. To this end, an extra cardinality consensus scheme may be performed \citep{Battistelli13,Li18CC}. That is to say, the fusion is performed in different means to the localization density and the cardinality distribution. 

 \subsection{Numerical Comparison between GM-AA and GM-GA}  
 
\begin{figure*}
\centering
\centerline{\includegraphics[width=14 cm]{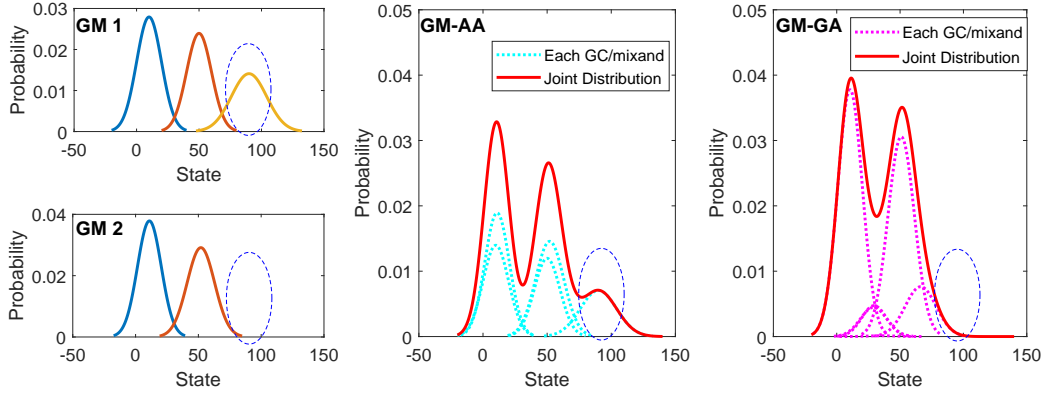}}
\caption{Unweighted AA and GA of two GMs consisting of components of different weights. The dashed ellipse indicates the position of a potential target. 
} \label{fig:PHD_fusion}
\end{figure*}

As addressed so far, the AA performs better in \textit{v}-fusion in the sense of yielding smaller bounds on the variance (cf. Remark 2) and MSE (cf. Remark 3) while the GA performs better in \textit{f}-fusion in the sense of always yielding smaller variance (cf. Remark 5) and smaller MSE in most cases (cf. Remarks 7 and 8). Therefore, we propose for the PHD fusion a hybrid means by using AA for cardinality fusion \citep{Li18CC} and GA for the localization density fusion. This is different from the usual, mere-GA GM fusion \citep{Battistelli13} 
, mere-AA GM fusion \citep{Li17gmMerging, Li17PCgm,Li17PCsmc,Li18FoV} or mere-cardinality fusion \citep{Li18CC}. Further, we note here that, in the case of labelled multi-target density where each label indicates a potential target, the fusion should match/associate the labels between the fusing sources first and then fuses the information (e.g., target existing probability, localization distribution) of the same target, not between different targets \citep{Li17PCgm}. 

To gain the insight into such a hybrid rules for unnormalized GM fusion, we consider an example in which two GMs are fused in the manner of unweighted AA fusion and unweighted GA fusion, respectively. Here, by unweighted we mean $\omega_1 = \omega_2 =\frac{1}{2}$. One GM referred to as GM 1 is given by three GCs (of weight sum 1.8) as follows 
$$f_1(x) = 0.7\mathcal{N}(x;10, 100)+ 0.6\mathcal{N}(x;50, 100)+ 0.5\mathcal{N}(x;90, 200),$$ 
and the other referred to as GM 2 is given by two GCs (of weight sum 1.7) as follows 
$$f_2(x) = 0.9\mathcal{N}(x;11, 100) + 0.8\mathcal{N}(x;52, 120).$$ 

As shown, the two GCs $\mathcal{N}(x;10, 100)$ and $\mathcal{N}(x;50, 100) $ in GM 1 match another two GCs $\mathcal{N}(x;11, 100)$ and  $\mathcal{N}(x;52, 120)$ in GM 2, respectively. They are likely indicating two respective targets. However, there is one extra GC $\mathcal{N}(x;90, 200)$ in GM 1, which could be either a false alarm (generated in GM 1) or a real detection (and then there is a misdetection in GM 2) - we hereafter refer to this GC as an isolated GC. The fusion results are given in Fig. \ref{fig:PHD_fusion} in which the fused GM-AA or GM-GA is given in the manner of showing each GC or showing the joint distribution of them, where the joint distribution is superimposition of those of each GC distribution along the state space. We obtain the following two remarks (the first of which is consistent with 
Remark 5):
 
\textbf{Remark 9}. \textit{The GA fusion generates more significant peaks and lighter tails than the AA fusion does.}
 
\textbf{Remark 10}. \textit{The isolated GC will survive (although its weighted will be reduced) in the AA fusion but will almost vanish in the GA fusion; this indicates that the GA fusion has better capability to suppress false alarm (if the isolated GC is a false alarm in practice) but will also suffer from misdetection (if the isolated GC turns out to be a real detection). This property is a double-edged sword. }
 
One more comment is in order. As we have addressed earlier in Section \ref{sec:background}.3, the support of the AA is the union of those of all initial functions while the support of the GA is the intersection of those of all initial functions. Therefore, assuming that both misdetection and false alarms are independent across fusing GMs, one complete misdetection occurred in one fusing GM (namely the support of the fusing distribution does not really cover the position of the misdetected target) will ``dominate'' the final GA result (namely the GA must suffer from the misdetection of that corresponding target), no matter how significant the detections are in the other fusing GMs and how many GMs there are. In fact, this problem becomes more serious when more fusing GMs are involved in the GA fusion because a missed detection at any fusing GM can degrade the GA result significantly, and the probability of such a missed detection in the GA obviously becomes larger when more sensors are involved; this may cause the counter-intuitive observation of the GA \textit{f}-fusion as more fusion leads to worse result \citep{Yu16, Li17PCsmc}. 

\begin{figure*}
\centering
\centerline{\includegraphics[width=17 cm]{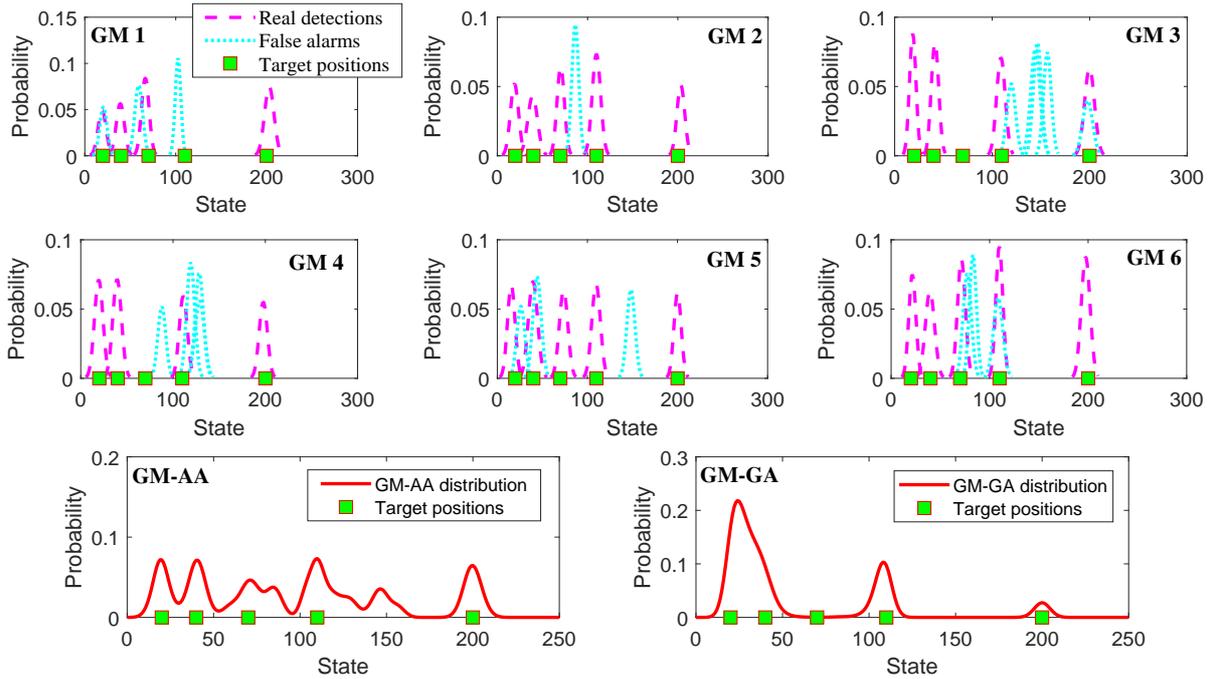}}
\caption{Unweighted AA and GA of six GMs consisting of both real detections and false alarms, both of which are given by weighted Gaussian distributions: the weights indicate the significance of the detections. The number of false alarms at each GM is Poisson distributed with rate 1 and the position of the false alarm is uniformly distributed in the interval between 0 and 200. There are also random misdetections in each GM as the detection probability is 0.9 for each target. } \label{fig:PHD_fusion_6}
\end{figure*}

To confirm the above analysis, we consider an example in which six GMs are fused, as shown in Fig. \ref{fig:PHD_fusion_6}. There are five targets in total which lie exactly at position 20, 40, 70, 110 and 200, respectively (as marked by green squares).  In each GM, any target is either detected with probability 0.9 and generates a detection (which are given in cyan print) or misdetected with probability 0.1. In addition, false alarms in each GM (which are marked in magenta print) are uniformly distributed in the interval [0, 200] and the number of false alarms is a Poisson random variable with rate 1. Fig. \ref{fig:PHD_fusion_6} shows the result for one trial based on the given statistics. In the result of the GA fusion, it is seen that the target lying at position 70 was mis-detected while the one lying at 200 was almost mis-detected (as only a low-weighted GC is generated). In the meanwhile, the detections of targets that lied at position 20 and 40 were mixed. These problems can be overcomed in the AA fusion which simply reserves all of the peaks (although this can be another problem). 
 
However, it is worth noting that the GA results we show here might be improved if numerical approximations such as those developed in \citep{Mariam07,Ahmed12,Gunay16,Li18Nehorai} were used in place of the analytic approximation \eqref{eq:GM_fractional_power}. 
 
Finally, we must stress that it is intractable to exactly compare between the results of the AA and of the GA for the multi-target density/PHD fusion in general due to the two fundamental issues that remain open. The first is regarding \textit{target state estimate extraction from the multi-target density/PHD} about a random (unknown) number of targets. Two of the most common solutions are referred to as Threshold and Rank rules, respectively. 
In the former, a threshold is specified in advance and the GCs/peaks whose associated weight is greater than that threshold will all be extracted as estimates while in the latter, the number of estimates is determined firstly and the corresponding number of GCs/peaks of the highest associated weights are extracted as the estimates. Both approaches have their own situation-sensitive strengths as well as deficiencies.
The second issue is regarding \textit{estimator evaluation metric} that has to trade-off between the penalty to misdetection and false alarm and the truth-to-detection distance, which highly depends on the practitioners' preference. 

\section{Conclusions} \label{sec:conclusion}
We have analyzed and compared the second order statistics of the GA and of the AA, in terms of averaging random variables (\textit{v}-fusion) or their PDFs (\textit{f}-fusion). 
The key findings that we have obtained can be summarized as follows:
\begin{itemize}
\item Given that all fusing estimators are unbiased, the AA is always unbiased while the GA may not be, in both \textit{v}-fusion and \textit{f}-fusion. Typically, when the fusing PDF is asymmetric, the GA tends to be biased (in the viewpoint of point estimation).  
\item For \textit{v}-fusion, 
\begin{enumerate}
    \item The variance of both AA and GA can be smaller than the smallest variance of the fusing variables given proper fusing weights, when the fusing variables are little or negatively correlated.
    \item For any two variables, the lowest AA variance (namely the lower bound) that can be yielded by adjusting the fusing weights is smaller than that of the GA variance. 
    \item The lowest bound of the MSE of either the AA or the GA is their corresponding variance, which is only possible when the real parameter lies between two variables and proper fusing weights are used. 
\end{enumerate}
\item For Gaussian \textit{f}-fusion, 
\begin{enumerate}
    \item The AA fusion always leads to a greater variance than the GA fusion does, for using the same fusion weights.
    \item The AA has an MSE that is the weighted average of the MSEs of the fusing estimators (and so it is bounded by the smallest and greatest MSEs).
    \item The GA fusion tends to perform better than the AA fusion in obtaining smaller MSE in most cases. 
\end{enumerate}
\item For \textit{PHD}-fusion based on a hybrid use of GA for distribution fusion and AA for cardinality fusion, 
 \begin{enumerate}
      \item The GA fusion generates more significant peaks and lighter tails than the AA fusion does; in order words, the GA is comparably more accurate but less robust.
      \item The GA fusion has better capability to suppress false alarm but also suffers from higher risk in causing misdetection as compared to the AA fusion, especially in the case of a large number of fusing sources. 
  \end{enumerate}
\end{itemize}

\section*{Appendix A: Lower Bound of $h(w)$ as in \eqref{eq:h(w)}}\label{sec:Appendix_A}
Here, we analyze the lower bound of function $h(w)$ as given in \eqref{eq:h(w)} for $w \in (0,1)$ and $\alpha\geq1, \rho\in(-1, 1)$.  Strightforwardly, the derivative of $h(w)$ with respect to $w$ is 
\begin{align} \label{eq:h(w)_derivative}
\frac{\mathrm{d} \hspace{0.5mm} h(w)}{\mathrm{d}w}
 = \big( 2(\alpha+1-2\rho \alpha^{\frac{1}{2}})w-2 + 2\rho \alpha^{\frac{1}{2}} \big).
\end{align}
Setting it to zero yields 
\begin{equation} \label{eq:mse_optimal_omega}
 w = \frac{1 -\rho \alpha^{\frac{1}{2}} }{1+\alpha-2\rho \alpha^{\frac{1}{2}}}, 
\end{equation}
This, however, may not satisfy the rule that $0<w<1$ and if not, cannot be used. We discuss two opposite cases:

\subsubsection*{1. When $\rho < \alpha^{-\frac{1}{2}}$:} 
In this case, \eqref{eq:mse_optimal_omega} satisfies $0<w<1$ and yields
\begin{equation} \label{eq:mse_f_bounds}
h(w)= \frac{
\alpha(1-\rho^2)
}
{
1+\alpha-2\rho \alpha^{\frac{1}{2}} 
}.
\end{equation}

Furthermore, by applying $\rho < \alpha^{-\frac{1}{2}}$, we obtain  $\frac{\mathrm{d}^2 h(w)}{\mathrm{d}w^2} 
 =  2(\alpha+1-2\rho \alpha^{\frac{1}{2}}) > 0$. This indicates that the bound given in \eqref{eq:mse_f_bounds} is indeed the lower bound. That is, if $\rho < \alpha^{-\frac{1}{2}}$, the optimal fusing weights to get the minimal $h(w)$ are given by 
\begin{equation} \label{eq:aa_optimal_omega}
\omega_1 = \frac{\alpha-\rho \alpha^{\frac{1}{2}} }{1+ \alpha-2\rho \alpha^{\frac{1}{2}}}, \hspace{2mm} \omega_2 = \frac{1 -\rho \alpha^{\frac{1}{2}} }{1+\alpha-2\rho \alpha^{\frac{1}{2}}}. 
\end{equation}

\subsubsection*{2. When $\rho \geq \alpha^{-\frac{1}{2}}$:} 
In this case, $w<0$ and so, \eqref{eq:mse_optimal_omega} can not be used. 
Considering that $h(w)$ is a convex function of $w$ and $\frac{\mathrm{d}^2 h(w)}{\mathrm{d}w^2} >0$, we obtain dual bounds of $h(w)$ at the boundaries of the support interval of the fusing weights, namely 
$$
1 = h(0) < h(w) < h(1) = \alpha.
$$

\section*{Appendix B: An example for averaging two functions }\label{sec:Appendix_B}

Supposing the real parameter $\theta = \frac{2 \hspace{0.5mm}\sqrt[]{2}}{3}$, an unbiased estimator 
is given with a uniform PDF on the interval $(0,1]$
, namely
$$
f_{\hat{\theta}_1}(x) = \left\{
\begin{array}{ll} 
\frac{3 \hspace{0.5mm}\sqrt[]{2}}{8} & \mathrm{if} \hspace{1mm} x\in(0,\frac{4 \hspace{0.5mm}\sqrt[]{2}}{3}] \hspace{0.5mm}, \\
0 & \mathrm{Otherwise}  \hspace{0.5mm}
.
\end{array} \right.$$
and another unbiased estimator is given with asymmetric PDF
$$
f_{\hat{\theta}_2}(x) = \left\{
\begin{array}{ll} 
x & \mathrm{if} \hspace{1mm} x\in(0, \hspace{0.5mm}\sqrt[]{2}] \hspace{0.5mm}, \\
0 & \mathrm{Otherwise}  \hspace{0.5mm}
.
\end{array} \right.$$

Simply for $\omega_1= \omega_2 =\frac{1}{2}$, the AA remains unbiased as \eqref{eq:unbiased_AA_f} holds but the GA \eqref{eq:GA-f} is 
$$
f_{\hat{\theta}_\mathrm{GA}}(x)=C^{-1} f_1^{\frac{1}{2}}f_2^{\frac{1}{2}}= \left\{
\begin{array}{ll} 
\frac{3}{2\sqrt[4]{8}}x^{\frac{1}{2}} & \mathrm{if} \hspace{1mm} x\in(0, \hspace{0.5mm}\sqrt[]{2}] \hspace{0.5mm}, \\
0 & \mathrm{Otherwise}  \hspace{0.5mm}
.
\end{array} \right.$$

Then, \eqref{eq:biased_GA_f} reads
$$
\bar{\theta}_f^{\mathrm{GA}} = \int_\mathcal{\Theta} xf_{\hat{\theta}_\mathrm{GA}}(x) dx 
= \frac{3 \hspace{0.5mm}\sqrt[]{2}}{5} \neq \theta \hspace{0.5mm}.$$


\section*{Acknowledgement}
This work was supported in part by the Marie Sk\l{}odowska-Curie Individual Fellowship under Grant 709267, in part by the Northwestern Polytechnical University and in part by the MOVIURBAN Project (Ref. SA070U 16) co-financed with Junta de Castilla y Le\'on, Consejer\'ia de Educacin and FEDER funds.







\bibliographystyle{model1-num-names} 
\bibliography{AAvsGA.bib}

\end{document}